\documentclass{pasj00}

 \SetRunningHead{H.~Imai \etal}
{VERA astrometry of H$_{2}$O masers in \i18286}
 \Received{2012/04/28}
 \Accepted{2012/10/16}
 \def \vlsr{$V_{\mbox{\scriptsize LSR}}$}
 \def \kms{~km~s$^{-1}$}
 \def \etal{~et~al.}
 
 \def \h2o{H$_{2}$O}
 \def \i18286{IRAS~18286$-$0959}

 \title
{Annual Parallax Distance and Secular Motion of the {\it Water Fountain} Source \i18286}

 \author{Hiroshi  \textsc{Imai}\altaffilmark{1,2}, Tomoharu \textsc{Kurayama}\altaffilmark{3}, 
Mareki \textsc{Honma}\altaffilmark{4}, and Takeshi \textsc{Miyaji}\altaffilmark{4}}

 \altaffiltext{1}{Department of Physics and Astronomy, Graduate School of Science and Engineering, \\
 Kagoshima University, 
1-21-35 Korimoto, Kagoshima 890-0065}
 \email{hiroimai@sci.kagoshima-u.ac.jp}

\altaffiltext{2}{International Centre for Radio Astronomy Research, M468, \\
The University of Western Australia, 35 Stirling Hwy, Crawley, Western Australia, 6009, Australia} 

 \altaffiltext{3}{Center for Fundamental Education, Teikyo University of Science, \\
2525 Yatsusawa, Uenohara, Yamanashi 409-0193}

 \altaffiltext{4}{Mizusawa VLBI Observatory, National Astronomical Observatory of Japan, \\
2-21-1 Osawa, Mitaka, Tokyo 181-8588}


\begin{document}

\maketitle


 \begin{abstract}
We report on results of astrometric observations of \h2o\ masers in the ``water fountain" source  \i18286 (I18286) with the VLBI Exploration of Radio Astrometry (VERA). These observations yielded an annual parallax of \i18286, $\pi=0.277\pm 0.041$~mas, corresponding to a heliocentric distance of $D=3.61^{+0.63}_{-0.47}$~kpc. The maser feature, whose annual parallax was measured, showed the absolute proper motion of $(\mu_{\alpha}, \mu_{\delta})=(-3.2\pm 0.3, -7.2 \pm 0.2)$[mas~yr$^{-1}$]. The intrinsic motion of the maser feature in the internal motions of the cluster of features in I18286 does not seem to trace the motion of the bipolar jet of I18286. Taking into account this intrinsic motion, the derived motion of the maser feature is roughly equal to that of the maser source I18286 itself. 
The proximity of I18286 to the Galactic midplane ($z\approx$10~pc) suggests that the parental star of the water fountain source in I18286 should be intermediate-mass AGB/post-AGB star, but the origin of a large deviation of the systemic source motion from that expected from the Galactic rotation curve is still unclear. 
\end{abstract}


\section{Introduction}

Energetic mass loss (rate up to $10^{-3}\dot{M_{\odot}}{\rm yr}^{-1}$) has been observed from dying stars such as asymptotic giant branch (AGB) and post-AGB stars. The spatio-kinematics of the mass-loss flows have been investigated in great detail by observations of maser emission such as SiO, \h2o, and OH using very long baseline interferometry (VLBI). These maser sources are associated with circumstellar envelopes (CSEs) of AGB stars, but in rare cases with highly collimated bipolar jets, so called ``water fountains," and CSE remnants around post-AGB stars or central objects of planetary nebulae (PNe). In the latest decade, high precision astrometry in very long baseline interferomery (VLBI) has enabled the measurement of trigonometric parallax distances and proper motions of maser sources. Even in the rare cases mentioned above, the heliocentric distances and three-dimensional secular motions of the water fountains, pre-PNe, and PNe have been measured, in those cases where \h2o\ masers could be detected \citep{ima07c,ima11b,taf11}. These kinematic approaches have contributed to estimating the physical parameters of these parental stars and their evolutionary properties. Taking into account their locations and 3D motions in the Milky Way, it has been suggested that these stars should be intermediate-mass AGB/post-AGB stars (e.g., \cite{ima07a}). 

Here we report on the measurement of an annual parallax of \h2o\ masers in \i18286 (hereafter abbreviated as I18286) with the VLBI Exploration of Radio Astrometry (VERA). The VERA astrometry for I18286 was conducted in one of the VERA key science projects, which focuses on \h2o\ maser sources at intermediate distances (2--5 kpc from the Sun) in order to extend the distance scale for the VERA parallax measurements. The result of the I18286 astrometry could  be used as one of the data points for exploring the Galactic dynamics if the motion of I18286 closely follows that of the Galactic rotation (i.e. \cite{rei09,hon12}). I18286 is a member of the class of water fountains, whose spatio-kinematics of \h2o\ masers have been investigated in detail. \citet{yun11} suggest that most of the \h2o\ masers in I18286 are associated with ``double helices" of highly collimated bipolar jets with speeds of $\sim$100\kms. The formation of the double helices is expected from discontinuous mass ejection from a moving star on a period of $\sim$30~yr. The annual parallax measurement for I18286 is the second case of parallax measurements for the water fountains after that for IRAS 19134$+$2131 \citep{ima07c}. This paper discusses the galactic kinematical properties of I18286 in Sect. \ref{sec:astrometry}. However, the existence of a low-velocity ``equatorial flow" should be taken into account when one interprets the observed secular motion of the maser feature in the cluster of features in I18286 \citep{ima07a}. The argument of the equatorial flow will be described in a separate paper in more detail.  


\section{Observations and data reduction}
\label{sec:observations}
The VERA observations of the I18286 \h2o\ ($J_{K_{-}K_{+}}=6_{12}\rightarrow 5_{23}$, 22.235080~GHz) masers were conducted at 16 epochs from 2007 October to 2009 September. Table \ref{tab:status} gives a summary of these observations, maser source mapping, and astrometry. Out of 16 epochs, 14 produced scientifically meaningful output. At each epoch, the observation was made for 6.7 hr in total. I18286 was observed together with the fringe-phase and position reference source, ICRF J183220.8$-$103511 (hereafter abbreviated as J1832), separated by 0\arcdeg.67 from I18286, simultaneously using VERA's dual-beam system. They were tracked for 30--35 min out of every 40 min, followed by scans on other band-pass calibrator sources. The received signals were digitized in four quantization levels, then divided into 16 base band channels (BBCs) with a bandwidth of 16 MHz each. One of the BBCs collected data from I18286 at the frequency band including the \h2o\ maser emission, while other BBCs from J1832 spanning a total frequency band range of 480~MHz. The BBC outputs had a recording data rate of 1024~Mbits~s$^{-1}$. The data correlation was made with the Mitaka FX correlator with a correlation accumulation period of 1~s. The correlation outputs consisted of 512 and 32 spectral channels for the \h2o\ maser and reference continuum emission, respectively. The former corresponds to a velocity spacing of 0.42\kms, which is narrow enough to resolve an \h2o {\it maser feature} (corresponding to a maser gas clump), which consists of two or more spectral channel components called {\it maser spots}. 

Data reduction was mainly made with the National Radio Astronomy Observatory (NRAO) Astronomical Image Processing System (AIPS) package. For astrometry, we need special procedures described as follows (see also e.g., \cite{hon07,ima07b}). Firstly, delay-tracking was repeated for the correlated data using better delay-tracking models calculated with the software equivalent to the CALC9 package developed by the Goddard Space Flight Center/NASA VLBI group. Throughout the whole data analysis, we adopted the coordinates of the delay-tracking center: 
$\alpha_{J2000}=$18$^{\mbox h}$31$^{\mbox m}$22$^{\mbox s}$\hspace{-2pt}.934,  
$\delta_{J2000}=$~$-$09$^{\circ}$57$^{\prime}$21\arcsec.70 for I18286 and 
$\alpha_{J2000}=$18$^{\mbox h}$32$^{\mbox m}$20$^{\mbox s}$\hspace{-2pt}.836521,  
$\delta_{J2000}=$~$-$10$^{\circ}$35$^{\prime}$11\arcsec.20055 for the reference source J1832. 
The delay-tracking solutions include residual delay contributions from the atmosphere, which were estimated using the global positioning system (GPS) data \citep{hon08b}. 
Secondly, differences in instrumental delays between two signal paths in the dual beam system were
calibrated using the differential delays, which were measured using artificial noise signals injected into the two receivers at the same time. The measurement accuracy has been improved for observations since 2007 by installing one to four artificial noise sources \citep{hon08a}. Thirdly, fringe-fitting and self-calibration were performed using the continuum source data, whose solutions were applied in the data analysis of maser emission. Only the solutions in the BBC at the same frequency as that for the \h2o\ maser emission were valid for the fringe-phase calibration. The accuracy of the phase-fluctuation compensation and the coherence of the integration is dependent on weather condition, particularly the humidity, which is seasonally variable. These effects affect the final astrometric accuracy. Finally, image cubes of the maser source were obtained by deconvolution through the CLEAN algorithm with a typical synthesized beam of 0.9$\times$2.5 in milliarcseconds (mas) in the case of full operation of VERA's four antennas. Because the CLEAN deconvolution was performed by automatically selecting local peaks of brightness in each of channel maps without smaller CLEAN boxes, side lobes of the synthesized beam pattern affected the final CLEAN image cube\footnote
{A tool for image deconvolution with automatic CLEAN box setting is currently in construction and will be used in future image synthesis. This tool works in a  Python/ParselTongue script. ParselTongue is a library developed in JIVE and can control AIPS tasks from the outside of AIPS POPS using Python scripts. See this URL. http://www.jive.nl/dokuwiki/doku.php?id=parseltongue:parseltongue}.  
Each maser spot (or velocity component) was identified as a Gaussian brightness component using the AIPS task SAD. 

In order to obtain higher quality maser maps, we also adopted a normal procedures for maser source mapping. Fringe-fitting and self-calibration were performed using a spectral channel that contain bright maser emission. Column 4 in table \ref{tab:status} gives the local-standard-of-rest (LSR) velocity of the spectral channel selected as phase- and position-reference. The obtained solutions of calibration were applied to the data in all spectral channels.

\section{Results}
\label{sec:results}

\subsection{The distribution and proper motions of \h2o maser features in \i18286}
\label{sec:proper-motions}

\begin{figure*}
\FigureFile(150mm,240mm){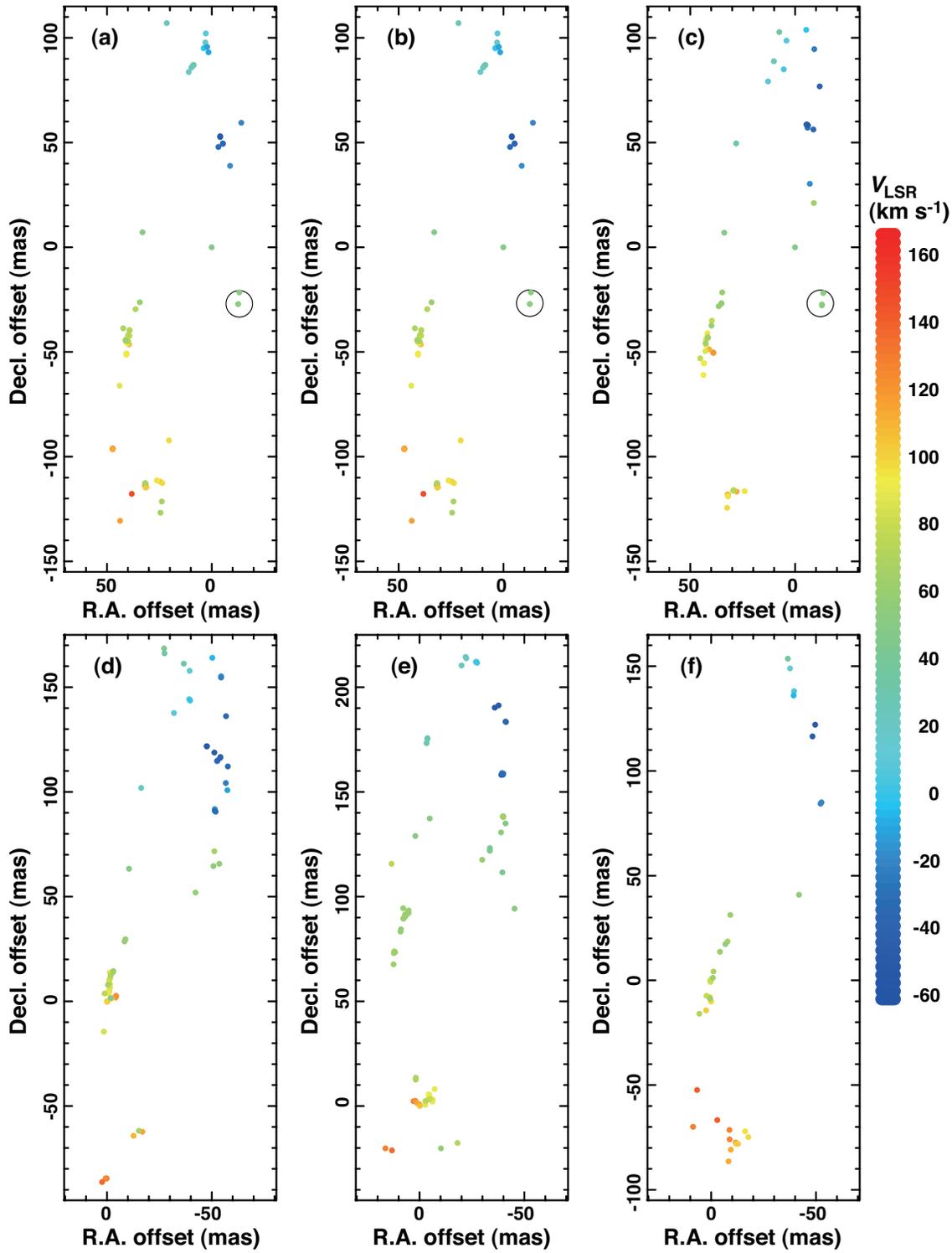}
\caption{Distributions of \h2o\ maser spots in \i18286. A black opened circle indicates the location of the maser feature (\i18286:I2013-{\it 14}) that contains maser spots measured their annual parallax. (a) On 2007 October 23. (b) On 2008 February 17. (c) On 2008 August 12. (d) On 2009 January 19. (e) On 2009 May 11. (f) On 2009 September 9. We note that some maser spots that were detected in the image cube obtained through phase-referencing calibration are missing in these plots, especially on 2009 January 19. Such maser spots include those detected in the phase-referencing image synthesis and measured their annual parallax.}
\label{fig:spots}
\end{figure*}

Column 6 of table \ref{tab:status} gives the numbers of maser features identified in the individual epochs.  Figure \ref{fig:spots} shows the distributions of \h2o\ maser features at six out of 14 epochs. The maser feature distribution was highly variable, with only a small fraction of the maser features having lifetimes longer than half year. Note that the VLBA observations conducted by \citet{yun11} and the VERA observations reported in this paper, respectively, lasted for similar seasons, 2008 April--2009 May and 2007 October--2009 September, respectively. The latter observations identified most of the same maser features as found in \citet{yun11}. However, there exists some discrepancy of maser feature detections between these observations because of the different image sensitivities\footnote
{The \h2o maser features of I18286 were resolved out in longer baselines of the VLBA as described in table 1 of \citet{yun11}. Also the total integration time of I18286 was shorter in the VLBA observations than that of the VERA observations. Therefore, the difference of the array sensitivity was not so significant as expected from full operation of VLBA's 10 antennas.} and, more importantly,  different epochs of observations. Although our VERA observations could not trace the maser feature that was selected as position-reference in the VLBA observations (at an LSR velocity \vlsr$\simeq$ 52\kms\ in the eastern side of the maser distribution) over a full year, we newly identify another maser feature that survived over one year in the western side of the maser distribution. This maser feature (\i18286:I2013-{\it 14}, see table \ref{tab:pmotions}), has a similar LSR velocity and is denoted by a black opened circle in figure \ref{fig:spots}. The maser spots included in this maser feature was used for the annual parallax measurement as described in subsection \ref{sec:astrometry}. 

\begin{figure*}
\begin{center}
\FigureFile(120mm,180mm){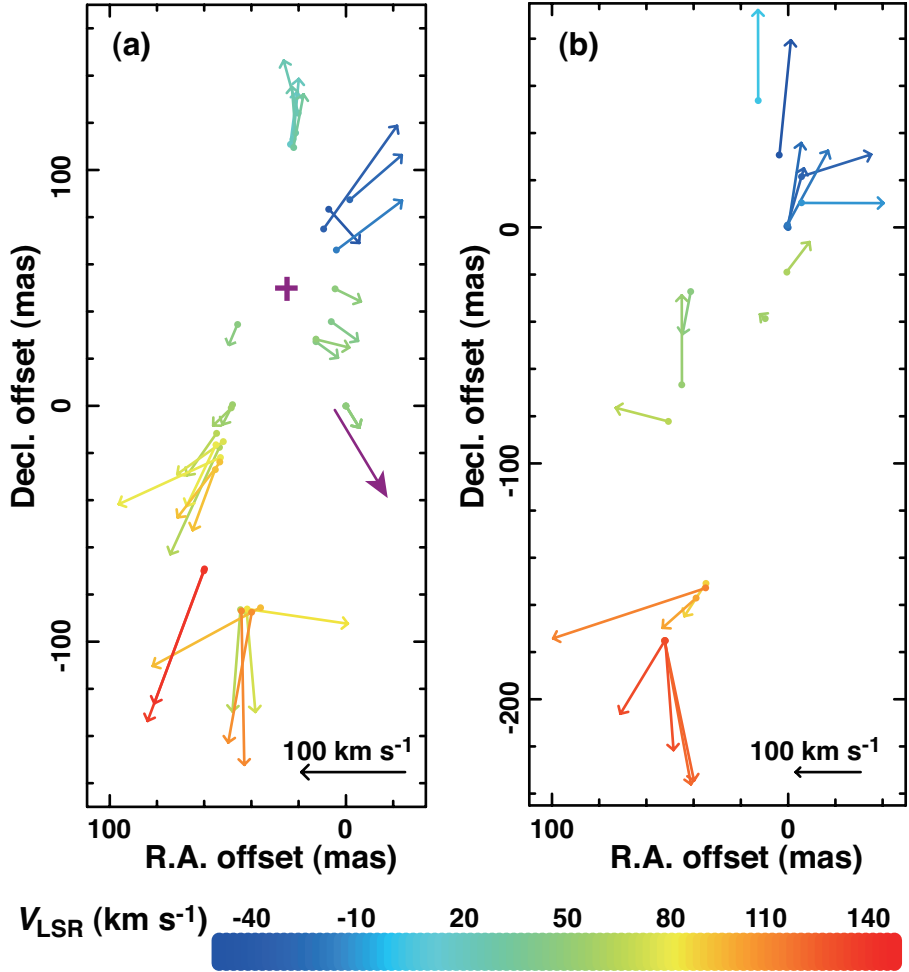}
\end{center}
\caption{Relative proper motions of \h2o\ maser spots in \i18286. The origin of coordinates is set to the position-reference maser feature, which was selected from different features between the epochs in 2007--2008 (\i18286:I2013-{\it 14}) and those in 2009 (\i18286:I2013-{\it 36}) . Colors of maser feature indicate LSR velocities. An arrow shows the relative proper motion of the maser feature. The root position of an arrow indicates the location of the maser feature at the first of the epochs when the feature was detected. The length and the direction of an arrow indicate the speed and direction of the maser proper motion, respectively.  (a) Proper motions identified in 2007--2008. The annual parallax and secular motions of maser spots were measured for those in the position-reference feature at the map origin. The systemic motion (or the motion of star), which was estimated in the least-squares model fitting, is subtracted from the measured proper motions. A magenta plus symbol denotes the estimated location of the dynamical center of the modelled outflow. A magenta arrow indicates a proper motion ($-51$, $-85$)[km~s$^{-1}$], corresponding to ($-3,-5$)[mas~yr$^{-1}$]. If the relative motion of the position-reference feature has this motion vector with respect to the system, the systemic motion well follows a motion expected from the Galactic rotation curve. (b) Same as (a) but in 2009. A mean motion of (20, $-130$) [km~s$^{-1}$] is subtracted from the individual measured motions.}
\label{fig:proper-motions}
\end{figure*}

The relative proper motions of maser features with respect to the position-reference feature were identified for the features detected at three or more epochs. Some proper motions were also removed from consideration because their LSR velocity drifts were higher than 10\kms yr$^{-1}$. Thus only 50 maser proper motions were measured although there existed a few hundred identified features. Because there was no single maser feature that was detected at all epochs, different maser features were selected as position reference in the first nine and final five epochs. Table \ref{tab:pmotions} gives the parameters of the maser features whose proper motions were measured. Figure \ref{fig:proper-motions} shows the spatial distribution of the relative maser proper motions. The spatio-kinematical structure roughly reproduces a fast bipolar outflow found by \citet{yun11}, but it is difficult to recognize the double helix jets as seen by \citet{yun11}. The VLBA identification of maser proper motions seems to have biases to high velocity motions. It is noteworthy that, in the present result, there exist a group of maser features moving in slow velocities ($V\lesssim$30\kms) near the center of the maser feature distribution. They were moving perpendicularly to the high velocity  ($V\gtrsim$100\kms) motions of maser features in the north--south direction. This implies the existence of an ``equatorial flow" \citep{ima07a}, different from the high velocity flows ($V\sim$100\kms) that are associated with highly collimated bipolar jets described by \citet{yun11}. This equatorial flow will be described in a separate paper. 

Nevertheless it is still necessary to estimate the motion of the star itself in I18286 (or the systemic motion) with respect to the maser proper motion in order to discuss the systemic motion of I18286 in the Milky Way (see subsection \ref{sec:discussion}). The maser motions shown in figure \ref{fig:proper-motions} are well approximated to an outflow with a point-symmetric velocity field. We performed the least-squares fitting method for the maser motion data using a radially expanding flow model. This method has been repeatedly described in our previous papers (e.g. \cite{ima11a,ima13}). Note that the model fitting assumes independent radial-expansion velocities of the maser features rather than a common radial velocity field as a function of distance from the dyamical center of the flow. This is, however,  sufficient for estimating the systemic motion of the outflow origin as one of the free parameters. Table \ref{tab:kinematic-model} gives the derived fitting parameters. Figure \ref{fig:proper-motions}a displays the measured maser motion vectors after subtracting the derived systemic motion vector.  The model-fitting result suggests that position-reference feature (\i18286:I2013-{\rm 14}) itself may belong to the equatorial flow mentioned above.

\subsection{The annual parallax distance to  \i18286}
\label{sec:astrometry}

\begin{figure*}
\FigureFile(170mm,150mm){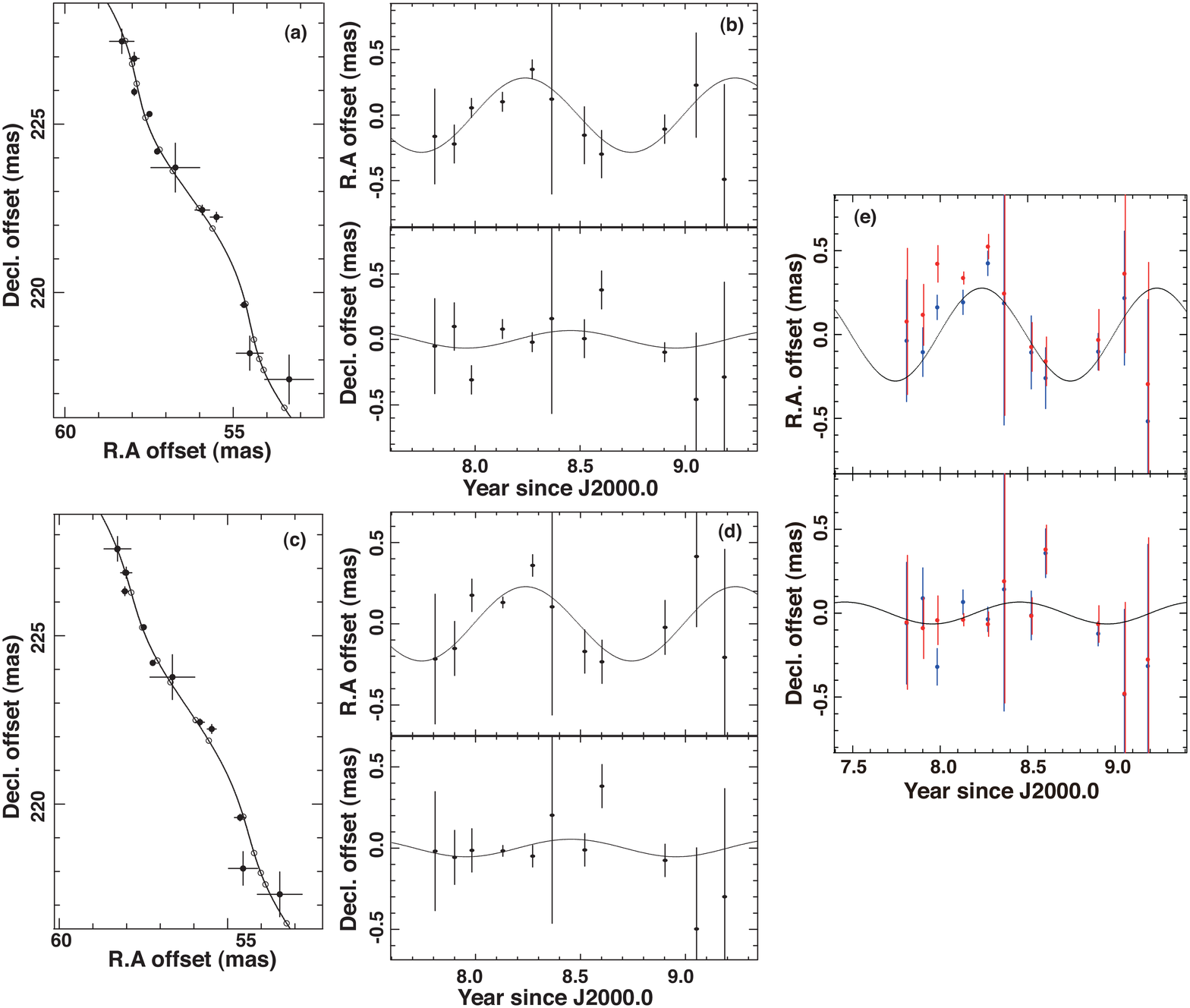}
\caption{Motions of the 53.4\kms\ and 53.0\kms\ components of \h2o masers in \i18286 and the kinematical models for these motions. (a) R.A. and decl. offsets with respect to the phase-tracking 
center of the 53.4\kms\ component. A filled circle shows the data point observed and used for the annual parallax measurement. A solid curve shows the modeled motion including an annual parallax and a constant velocity proper motion. An opened circle indicates the spot position expected in the model at the observation epoch. (b) R.A. variation of the 53.0\kms\ component along time. 
The estimated linear proper motion is subtracted from the observed spot position.  
A solid curve shows the modeled annual parallactic motion.  
(c) Same as (a) but for the 53.4\kms\ component. 
(d) Same as (b) but for the 53.0\kms\ component. 
(e) The result of the combined annual parallax fitting. Blue and red data points shows those of the 53.4\kms\ and 53.0\kms\ components, respectively. For clarity, data points are slightly shifted in the horizontal axis.}
\label{fig:model-motions}
\end{figure*}

Column 7 of table \ref{tab:status} shows validity of astrometry for the annual parallax measurement in the individual epochs
(Y/N/S). As described in subsection \ref{sec:proper-motions} and table \ref{tab:status}, the maser feature \i18286:I2013-{\it 14} was identified in the observations between 2007 October and 2009 March in the maser image cubes synthesized through self-calibration and/or phase-referencing procedures and two maser spots in this feature were used for the parallax measurement. Figure \ref{fig:model-motions} shows the motions of these maser spots. The spot motions can be fitted to combination of an annual parallax and a constant-velocity motion as described later. The uncertainty of spot position presented in figure \ref{fig:model-motions} indicates only the contribution from thermal noise, which is underestimated if the spot is extended. We note that \citet{yun11} and the data of \citet{ima07a} suggest that the \h2o maser spots in I18286 were extended. Thus, deviations of the data points from the modeled motions (up to 0.5~mas) are larger than the position errors expected from combination of contributions from thermal noise, instrumental, and atmospheric phase-delay residuals (0.1~mas level, \cite{hon10}, see also general formulation in \cite{asa07}). They should be mainly attributed to the variation of maser spot structures. Note that the temporal variation of maser spot/feature structure is not necessarily random as assumed in \citet{hon10} if the feature is associated with some physical feature (e.g. a shock front, \cite{ima02}) changing on a specific time scale. Nevertheless, random variation of a spot position around its ballistic motion as shown in figure \ref{fig:model-motions} is expected when the maser is highly spatially resolved with a VLBI synthesized beam whose shape may change from one epoch to another as small intrinsic variation of the maser structure will be highly enhanced (c.f. \cite{ima07b}). 

We attempted the least-square method for the spot motions to fit the modeled motions each of which is composed of an annual parallax, a constant secular motion, and a position offset at the reference epoch (J2000.0)\footnote
{In the model fitting, all spot positions are at first converted to the relative positions with respect to that at the first observation epoch for conserving the fitting accuracy.}. In order to obtain the common annual parallax of these spots, iterative procedures were adopted, which is similar to the approach by \citet{san12}. Firstly, the model fitting was performed independently for the individual spot motions (independent fitting). Secondly, a mean annual parallactic motion was subtracted from the spot motions to estimate only the position offsets and secular motions of spots (proper motion fitting). Thirdly, the systemic motions estimated from the derived parameters were subtracted from the original spot motions. Finally, the position residuals were used to estimate a common parallax and position offset and proper motion residuals (combined fitting). The proper motion fitting was again performed using the estimated common parallax. Then the procedures from the proper motion to combined fitting were iterated until the estimated parameters seemed converged. 

Table \ref{tab:astrometry} gives the derived parameters after the final iteration. Although the two maser spots were associated with the same maser feature and their motions are not completely independent, the data combination may mitigate the error contribution from statistical (large) errors expected from the signal-to-noise ratios of maser spot detection and the random variation of spot structures. In the model fitting, weighting with position accuracy was adopted as performed in our previous analysis \citep{ima11b}. Thus we obtained an annual parallax of I18286, $\pi=0.277\pm 0.041$~mas, corresponding to a distance value of $D=3.61^{+0.63}_{-0.47}$~kpc. The kinematic distance to I18286 of 3.1~kpc \citep{deg07} is roughly consistent  with the annual parallax distance.

The {\it absolute proper motion} of the maser feature \i18286:I2013-{\it 14} was derived to be $(\mu_{\alpha}, \mu_{\delta})=(-3.2\pm 0.3, -7.2 \pm 0.2)$[mas~yr$^{-1}$] from the mean motions of the two spots, whose parameters are given in table \ref{tab:astrometry}. Here we estimate the location and the three-dimensional motion of the I18286 system in the Milky Way. Table \ref{tab:MW-motion} gives the derived parameters. I18286 is located very close to the Galactic midplane ($z\simeq$7~pc). On the other hand, the systemic secular motion of I18286 has some uncertainty, described as follows, which is dependent on the relative motion of the feature \i18286:I2013-{\it 14} with respect to the central star. If the motion of this maser feature is the same as that of the star (Case 1 in table \ref{tab:MW-motion}), the systemic motion has a very large deviation (up to 100\kms) from that expected from the Galactic rotation curve. This is a similar case to that in which one adopts the relative motion of the feature estimated from the model fitting as described in subsection \ref{sec:proper-motions} (Case 2). The systemic motion only follows the Galactic rotation curve closely if one assumes a relative proper motion of the feature ($-3,-5$) [mas~yr$^{-1}$], or ($-51$, $-85$) [km~s$^{-1}$] with respect to the star, as indicated with a magenta arrow in figure \ref{fig:proper-motions} (Case 3).

\section{Discussion}
\label{sec:discussion}

Through the present work, we can obtain some constraint on the properties of the central star in the water fountain I18286. The annual parallax distance to I18286 gives the luminosity value of I18286, $\sim 1.2\times 10^4\; L_{\odot}$, a factor of 1.4 higher than the value previously derived ($8.7\times 10^3\;L_{\odot}$, \cite{deg07}). This suggests that I18286 has a luminosity higher than the typical value of AGB stars ($\sim$6 000~$L_{\odot}$), supporting the hypothesis of I18286 to be a higher mass AGB/post-AGB star  (e.g., \cite{ima07a, ima07c}). Note that the high mass population of evolved stars such as red supergiants follows closely the Galactic rotation within 30\kms (e.g. \cite{rei09, asa10, ima13}) while lower mass evolved stars may have the large kinematical deviations due to the dynamical relaxation in the Milky Way during their long lifetime. In the case of I18286, the large deviation of its motion from the Galactic rotation favors the hypothesis of an AGB star over that of a red supergiant. However, its close proximitiy to the Galactic mid-plane does not rule out the possibility that the large kinematical deviation is attributed to a binary motion in I18286. SiO maser or millimeter continuum emission can pinpoint the location of the star \citep{ima05,ima07a}, but has not yet been detected toward I18286. Detection of a circumstellar envelope in molecular line emission with a new facility such as the Atacama Large Millimeter/submillimeter Array (ALMA) will shed light on the property of of the central star. In practice, it is difficult to distinguish the intrinsic emission of I18286 from interstellar foreground/background emission \citep{ima09} although this object is clearly identified in the {\it Spitzer}/GLIMPSE image \citep{deg07}. In addition, more intensive monitoring VLBI observations of the I18286 \h2o masers are necessary in order to trace a greater number of proper motions of maser features including shorter-lived (a few months) features, enabling to elucidate the maser spatio-kinematics in more detail, including a possible equatorial flow or an AGB envelope. 

\bigskip
VERA/Mizusawa VLBI observatory is a branch of the National Astronomical Observatory of Japan, an interuniversity research institute supported by the Ministry of Education, Culture, Sports, Science and Technology. 
We acknowledge all staff members and students who have helped in array operation and in data correlation of the VERA. We also thank Richard Dodson for carefully reading the manuscript and give us useful comments. HI was financially supported by Grant-in-Aid for Young Scientists from the Ministry of Education, Culture, Sports, Science, and Technology (18740109) as well as by Grant-in-Aid for Scientific Research from Japan Society for Promotion Science (JSPS) (20540234 and 22-00022). HI also acknowledge for the support for his stay at ICRAR in the Strategic Young Researcher Overseas Visits Program for Accelerating Brain Circulation funded by JSPS.

\clearpage
\begin{table*}[h]
\caption{Parameters of the VERA observations}\label{tab:status}

\scriptsize
\begin{tabular}{l@{ }lc@{ }r@{ }cl@{ }rc@{ }c@{ }l@{}} \hline \hline
\multicolumn{2}{l}{\underline{\hspace{1cm}Observation\hspace{1cm}}}
& VERA & $V_{\rm ref}$\footnotemark[b] & & \multicolumn{1}{c}{Beam\footnotemark[d]} 
& & \multicolumn{3}{l}{\underline{\hspace*{2.7cm}Astrometry\hspace{2.7cm}}} \\
code & epoch  & telescopes\footnotemark[a] & (km~s$^{-1}$) & Noise\footnotemark[c] 
&  \multicolumn{1}{c}{[mas]} & $N_{\mbox{\scriptsize s}}$\footnotemark[e] 
& valid?\footnotemark[f] & $I_{\rm peak}$\footnotemark[g] & Note \\ \hline 
r07296a \dotfill & 2007 October 23 & MROS & 48.2 & 42
& 4.34$\times$0.98, $-47^{\circ}$\hspace{-2pt}.2 & 68 & Y & 1.5  & Calibration solutions invalid (O) \\
r07328a \dotfill & 2007 November 26 & MROS & 48.2 & 42  
& 2.35$\times$0.91, $-41^{\circ}$\hspace{-2pt}.1 & 71 & Y & 1.8  & \\
r07357b \dotfill & 2007 December 26 & MROS & 48.0 & 31  
& 2.64$\times$0.83, $-37^{\circ}$\hspace{-2pt}.1 & 60 & Y & 1.3 & \\ 
r08048b \dotfill & 2008 February 17 & MROS & 46.7 & 24  
& 1.63$\times$0.88, $-30^{\circ}$\hspace{-2pt}.2 & 105 & Y & 2.5 & \\ 
r08098a \dotfill & 2008 April 9 & MROS & 70.1 & 37  
& 1.65$\times$0.96, $-32^{\circ}$\hspace{-2pt}.5 & 73 & Y & 2.4 & \\ 
r08132a \dotfill & 2008 May 13 & MROS & 48.3 & 46
& 4.09$\times$0.90, $-37^{\circ}$\hspace{-2pt}.8 & 63 & S & 2.3 & See comment \footnotemark[f] \\ 
r08166a \dotfill & 2008 June 16 & (R)(O)S & --- & ---  
&  \multicolumn{1}{c}{---} & --- & N & & Earthquake (M), heavy rain (R) \\ 
r08174a \dotfill & 2008 June 24 & MOS & --- & --- 
&  \multicolumn{1}{c}{---} & --- & N & & High noise temperature (O) \\ 
r08189a \dotfill & 2008 July 9  & MROS & 86.3 & 90 
& 1.31$\times$1.24, $+65^{\circ}$\hspace{-2pt}.6 &  61 & Y & 2.8  & Too faint detected reference source \\ 
r08223a \dotfill & 2008 August 12 & MROS & 48.1 & 67
& 1.59$\times$1.18, $-49^{\circ}$\hspace{-2pt}.0 & 53 & Y & 2.5 & \\ 
r08329a \dotfill & 2008 December 26 & MROS & 83.7 & 50 
& 2.22$\times$1.05, $+89^{\circ}$\hspace{-2pt}.6 & 97 &  Y & 1.0 & \\ 
r09019b \dotfill & 2009 January 19 & MROS & 100.7 & 58 
& 4.06$\times$0.77, $-38^{\circ}$\hspace{-2pt}.2 & 69 
& Y & 1.0  &  See comment \footnotemark[h] \\
r09050b \dotfill & 2009 February 19 & M(R)O(S) & 103.5 & 100 
& 2.68$\times$0.95, $+14^{\circ}$\hspace{-2pt}.9 & 34 & N & & High noise temperatures (R, S) \\ 
r09068b \dotfill & 2009 March 9 & MROS & 94.2 & 40
& 2.17$\times$0.86, $-35^{\circ}$\hspace{-2pt}.9 & 104 & S & 0.6 & See comment \footnotemark[f]  \\ 
r09129b \dotfill & 2009 May 11 & MROS & 107.1 & 30 
& 2.46$\times$0.83, $-36^{\circ}$\hspace{-2pt}.1 & 74 & N & & \\ 
r09250a \dotfill & 2009 September 9 & MR(O)S & 82.7 & 89
& 2.36$\times$0.99, $-46^{\circ}$\hspace{-2pt}.0 & 39 & N & & High noise temperatures (O) \\
\hline
\end{tabular}

\footnotemark[a]Telescope whose data were valid for phase-referencing maser imaging. 
M: Mizusawa, R: Iriki, O: Ogasawara, S: Ishigakijima.
The station with parentheses had some problem during the observations and affected the annual parallax measurements. \\
\footnotemark[b]LSR velocity at the phase-referenced spectral channel. \\
\footnotemark[c]Smallest rms noise in the emission-free spectral channel image obtained through self-calibration image synthesis, in units of mJy beam$^{-1}$. \\ 
\footnotemark[d]Synthesized beam size resulting from natural weighted visibilities, i.e. major and minor axis lengths and position angle.\\
\footnotemark[e]Number of the detected maser features in the data obtained through self-calibration. \\
\footnotemark[f]Y and N: valid and invalid data point for annual parallax measurement, respectively. S: 
valid in a special procedure, in which the maser position was obtained from the relative position of another maser spot whose coordinates were measured in both of the image cubes synthesized through self-calibration and phase-referencing procedures. \\
\footnotemark[g]Peak intensity, in units of Jy~beam$^{-1}$, of the phase-referencing-based image of the maser feature  \i18286:I2013-{\it 14} that contained the maser spots selected for the astrometry. \\ 
\footnotemark[h]A too long fringe-fitting solution interval (3~min.) was adopted. The maser image cube obtained through self-calibration did not have better quality than that through phase-referencing calibration, causing some of the maser features found in the latter image cube missing in the former one. 
\end{table*}

\clearpage
\begin{table*}[h]
\caption{Parameters of the \h2o  maser features identified by 
proper motion toward \i18286} \label{tab:pmotions}
\scriptsize
\begin{tabular}{lrrrrrrrr    c@{ }c@{ }c@{ }c@{ }c@{ }c@{ }c@{ }c@{ }c } \hline \hline          
 & \multicolumn{2}{c}{Offset}
 & \multicolumn{4}{c}{Proper motion\footnotemark[b]}
 & \multicolumn{2}{c}{Radial motion\footnotemark[c]}
 & \multicolumn{9}{c}{Detection at} \\                                                                                                
 & \multicolumn{2}{c}{(mas)} 
 & \multicolumn{4}{c}{(mas yr$^{-1}$)}
 & \multicolumn{2}{c}{(km s$^{-1}$)}
 & \multicolumn{9}{c}{individual epochs} \\                                               
 & \multicolumn{2}{c}{\ \hrulefill \ } 
 & \multicolumn{4}{c}{\ \hrulefill \ } 
 & \multicolumn{2}{c}{\ \hrulefill \ } 
 & \multicolumn{9}{c}{\ \hrulefill \ } \\                                                           
 Feature\footnotemark[a] & $\Delta$R.A. & $\Delta$decl. & $\mu_{x}$ & $\sigma \mu_{x}$ & $\mu_{y}$ 
 & $\sigma \mu_{y}$ & V$_{z}$ & $\Delta$V$_{z}$\footnotemark[d] 
 & 1& 2& 3& 4& 5& 6& 7& 8& 9 \\ \hline                                                   
\hline
\multicolumn{18}{c}{Maser motions identified in 2007--2008 (9 epochs)} \\ \hline

  1   \ \dotfill \  &$     7.20$&$    83.39$&$  -0.95$&   0.48 &$  -0.68$&   0.35
 &$ -42.84$&   2.53
 &$\times$ &$\times$ &$\times$ &$\times$ &$\circ$ &$\times$ &$\circ$ &$\times$ &$\circ$   \\                   
  2   \ \dotfill \  &$     9.42$&$    75.00$&$  -3.38$&   0.33 &$   7.05$&   0.22
 &$ -30.99$&   1.19
 &$\circ$ &$\times$ &$\circ$ &$\circ$ &$\times$ &$\times$ &$\times$ &$\times$ &$\times$   \\                   
 3   \ \dotfill \  &$    -1.66$&$    87.27$&$  -2.17$&   0.33 &$   3.76$&   0.43
 &$ -25.55$&   1.83
 &$\times$ &$\times$ &$\circ$ &$\circ$ &$\circ$ &$\times$ &$\times$ &$\times$ &$\times$   \\                   
 4   \ \dotfill \  &$     3.91$&$    66.01$&$  -2.92$&   0.14 &$   4.02$&   0.17
 &$ -14.68$&   1.69
 &$\circ$ &$\circ$ &$\circ$ &$\circ$ &$\times$ &$\circ$ &$\circ$ &$\times$ &$\times$   \\                      
 5   \ \dotfill \  &$    23.54$&$   110.70$&$   0.38$&   0.24 &$   4.19$&   0.31
 &$  22.11$&   0.94
 &$\circ$ &$\circ$ &$\circ$ &$\circ$ &$\times$ &$\times$ &$\times$ &$\times$ &$\times$   \\                    
 6   \ \dotfill \  &$    22.33$&$   112.97$&$   0.45$&   0.23 &$   4.64$&   0.27
 &$  25.23$&   2.53
 &$\circ$ &$\circ$ &$\circ$ &$\circ$ &$\circ$ &$\times$ &$\times$ &$\times$ &$\times$   \\                     
 7   \ \dotfill \  &$    20.18$&$   124.11$&$   1.58$&   0.40 &$   4.19$&   0.55
 &$  29.46$&   0.77
 &$\times$ &$\times$ &$\circ$ &$\circ$ &$\circ$ &$\times$ &$\times$ &$\times$ &$\times$   \\                   
 8   \ \dotfill \  &$    21.35$&$   115.60$&$   0.96$&   0.75 &$   3.84$&   0.62
 &$  32.78$&   1.12
 &$\times$ &$\times$ &$\times$ &$\circ$ &$\circ$ &$\circ$ &$\times$ &$\times$ &$\times$   \\                   
 9   \ \dotfill \  &$    22.26$&$   109.60$&$   0.19$&   0.48 &$   4.22$&   0.64
 &$  38.72$&   2.95
 &$\times$ &$\times$ &$\times$ &$\circ$ &$\circ$ &$\circ$ &$\times$ &$\times$ &$\times$   \\                   
 10   \ \dotfill \  &$     6.18$&$    35.69$&$  -0.75$&   0.22 &$   0.14$&   0.23
 &$  46.69$&   2.11
 &$\times$ &$\times$ &$\times$ &$\circ$ &$\circ$ &$\circ$ &$\circ$ &$\times$ &$\times$   \\                    
 11   \ \dotfill \  &$    12.59$&$    27.09$&$  -0.47$&   0.08 &$   0.34$&   0.07
 &$  48.07$&   1.75
 &$\circ$ &$\circ$ &$\circ$ &$\circ$ &$\circ$ &$\circ$ &$\times$ &$\times$ &$\times$   \\                      
 12   \ \dotfill \  &$    45.96$&$    34.43$&$   1.26$&   0.08 &$   0.06$&   0.11
 &$  51.31$&   1.09
 &$\times$ &$\times$ &$\times$ &$\circ$ &$\circ$ &$\circ$ &$\circ$ &$\circ$ &$\times$   \\                     
 13   \ \dotfill \  &$     4.49$&$    49.52$&$  -0.68$&   0.35 &$   0.52$&   0.40
 &$  53.00$&   0.63
 &$\times$ &$\times$ &$\times$ &$\circ$ &$\circ$ &$\circ$ &$\times$ &$\times$ &$\times$   \\                   
 14   \ \dotfill \  &$     0.00$&$     0.00$&$   0.00$&   0.23 &$   0.00$&   0.36
 &$  53.33$&   1.69
 &$\circ$ &$\circ$ &$\circ$ &$\circ$ &$\circ$ &$\circ$  &$\circ$  &$\circ$ &$\circ$    \\                  
 15   \ \dotfill \  &$    12.64$&$    28.14$&$  -1.11$&   0.47 &$   0.77$&   0.44
 &$  56.00$&   0.84
 &$\times$ &$\times$ &$\circ$ &$\circ$ &$\times$ &$\circ$ &$\times$ &$\times$ &$\times$   \\                   
 16   \ \dotfill \  &$    47.95$&$     0.35$&$   1.34$&   0.17 &$   0.12$&   0.26
 &$  60.18$&   0.84
 &$\times$ &$\times$ &$\times$ &$\times$ &$\times$ &$\times$ &$\circ$ &$\circ$ &$\circ$   \\                   
 17   \ \dotfill \  &$    48.45$&$    -0.94$&$   1.80$&   0.21 &$   0.24$&   0.29
 &$  62.69$&   1.82
 &$\times$ &$\times$ &$\times$ &$\circ$ &$\times$ &$\circ$ &$\times$ &$\circ$ &$\times$   \\                   
 18   \ \dotfill \  &$    53.54$&$   -17.53$&$   3.53$&   0.31 &$  -4.82$&   0.47
 &$  67.34$&   1.06
 &$\circ$ &$\circ$ &$\circ$ &$\circ$ &$\times$ &$\times$ &$\times$ &$\times$ &$\times$   \\                    
 19   \ \dotfill \  &$    54.81$&$   -11.57$&$   2.43$&   0.60 &$  -1.16$&   0.81
 &$  70.98$&   0.63
 &$\circ$ &$\circ$ &$\circ$ &$\times$ &$\times$ &$\times$ &$\times$ &$\times$ &$\times$   \\                   
 20   \ \dotfill \  &$    44.67$&$   -86.43$&$   1.21$&   0.10 &$  -4.51$&   0.14
 &$  71.11$&   4.47
 &$\times$ &$\circ$ &$\circ$ &$\circ$ &$\circ$ &$\circ$ &$\times$ &$\times$ &$\times$   \\                     
 21   \ \dotfill \  &$    41.99$&$   -87.67$&$   0.28$&   0.48 &$  -4.38$&   0.54
 &$  79.20$&   2.95
 &$\times$ &$\times$ &$\times$ &$\times$ &$\times$ &$\circ$ &$\circ$ &$\circ$ &$\times$   \\                   
 22   \ \dotfill \  &$    51.93$&$   -15.20$&$   3.36$&   0.09 &$  -0.61$&   0.13
 &$  80.25$&   2.11
 &$\circ$ &$\times$ &$\circ$ &$\circ$ &$\circ$ &$\circ$ &$\times$ &$\times$ &$\times$   \\                     
 23   \ \dotfill \  &$    53.07$&$   -21.86$&$   6.51$&   0.24 &$  -1.41$&   0.20
 &$  84.80$&   0.28
 &$\times$ &$\times$ &$\times$ &$\times$ &$\circ$ &$\times$ &$\circ$ &$\circ$ &$\times$   \\                   
 24   \ \dotfill \  &$    55.07$&$   -16.65$&$   2.40$&   0.26 &$  -2.21$&   0.22
 &$  85.88$&   2.95
 &$\times$ &$\times$ &$\times$ &$\times$ &$\times$ &$\times$ &$\circ$ &$\circ$ &$\circ$   \\                   
 25   \ \dotfill \  &$    41.72$&$   -86.20$&$  -4.91$&   0.55 &$   0.40$&   0.49
 &$  89.71$&   1.54
 &$\times$ &$\times$ &$\circ$ &$\times$ &$\circ$ &$\circ$ &$\times$ &$\times$ &$\times$   \\                   
 26   \ \dotfill \  &$    55.42$&$   -27.10$&$   2.02$&   0.18 &$  -2.18$&   0.12
 &$  96.92$&   1.85
 &$\times$ &$\times$ &$\times$ &$\times$ &$\circ$ &$\circ$ &$\circ$ &$\circ$ &$\circ$   \\                      
 27   \ \dotfill \  &$    53.28$&$   -23.88$&$   3.16$&   0.57 &$  -1.89$&   0.46
 &$  96.94$&   4.00
 &$\circ$ &$\circ$ &$\circ$ &$\circ$ &$\times$ &$\times$ &$\times$ &$\times$ &$\times$   \\                    
 28   \ \dotfill \  &$    36.23$&$   -85.48$&$   6.83$&   0.19 &$  -2.07$&   0.26
 &$  99.64$&   0.92
 &$\circ$ &$\circ$ &$\circ$ &$\circ$ &$\circ$ &$\times$ &$\times$ &$\times$ &$\times$   \\                     
 29   \ \dotfill \  &$    39.85$&$   -87.57$&$   2.09$&   0.57 &$  -6.12$&   0.80
 &$ 113.63$&   1.40
 &$\times$ &$\times$ &$\times$ &$\times$ &$\times$ &$\circ$ &$\circ$ &$\circ$ &$\times$   \\                   
 30   \ \dotfill \  &$    44.25$&$   -86.89$&$   0.60$&   0.23 &$  -7.47$&   0.25
 &$ 115.88$&   3.05
 &$\circ$ &$\circ$ &$\circ$ &$\circ$ &$\circ$ &$\circ$ &$\times$ &$\times$ &$\times$   \\                      
 31   \ \dotfill \  &$    60.12$&$   -69.93$&$   3.94$&   0.27 &$  -7.25$&   0.42
 &$ 139.65$&   2.63
 &$\times$ &$\circ$ &$\circ$ &$\circ$ &$\circ$ &$\times$ &$\times$ &$\times$ &$\times$   \\                    
 32   \ \dotfill \  &$    59.92$&$   -69.14$&$   3.59$&   1.24 &$  -6.36$&   1.12
 &$ 144.29$&   4.21
 &$\circ$ &$\circ$ &$\circ$ &$\times$ &$\times$ &$\times$ &$\times$ &$\times$ &$\times$   \\                   
\hline
\multicolumn{18}{c}{Maser motions identified in 2009 (5 epochs)} \\ \hline
 33   \ \dotfill \  &$     3.69$&$    30.62$&$   0.13$&   0.13 &$   2.74$&   0.15
 &$ -48.16$&   3.37
 &$\times$ &$\times$ &$\circ$ &$\circ$ &$\circ$   \\                                                           
 34  \ \dotfill \  &$     0.45$&$     0.42$&$  -0.34$&   0.33 &$  -2.39$&   0.41
 &$ -30.30$&   1.83
 &$\circ$ &$\times$ &$\circ$ &$\circ$ &$\times$   \\                                                           
 35  \ \dotfill \  &$    -5.78$&$    21.67$&$  -5.08$&   0.77 &$  -5.64$&   1.27
 &$ -29.06$&   2.11
 &$\circ$ &$\circ$ &$\circ$ &$\times$ &$\times$   \\                                                           
  36  \ \dotfill \  &$     0.00$&$     0.00$&$   0.00$&   0.24 &$   0.00$&   0.33
 &$ -22.84$&   3.67
 &$\circ$ &$\circ$ &$\circ$ &$\circ$ &$\circ$   \\                                                             
 37  \ \dotfill \  &$     0.40$&$     1.00$&$  -2.50$&   0.39 &$  -0.93$&   0.32
 &$ -16.36$&   3.93
 &$\circ$ &$\times$ &$\circ$ &$\times$ &$\circ$   \\                                                           
 38   \ \dotfill \  &$    -5.71$&$    10.39$&$  -6.21$&   0.70 &$  -7.63$&   1.11
 &$  -9.38$&   2.18
 &$\circ$ &$\circ$ &$\circ$ &$\times$ &$\times$   \\                                                           
 39  \ \dotfill \  &$    12.68$&$    53.80$&$   1.19$&   0.35 &$   0.53$&   0.47
 &$   7.42$&   2.32
 &$\circ$ &$\circ$ &$\circ$ &$\circ$ &$\circ$   \\                                                             
 40  \ \dotfill \  &$    41.29$&$   -27.17$&$   1.91$&   0.56 &$ -11.46$&   0.44
 &$  50.99$&   1.96
 &$\circ$ &$\times$ &$\circ$ &$\circ$ &$\times$   \\                                                           
 41  \ \dotfill \  &$    44.96$&$   -66.61$&$   1.19$&   0.33 &$   0.42$&   0.38
 &$  57.01$&   0.84
 &$\times$ &$\times$ &$\circ$ &$\circ$ &$\circ$   \\                                                           
 42  \ \dotfill \  &$     0.64$&$   -18.83$&$  -0.89$&   0.29 &$  -4.94$&   0.28
 &$  66.95$&   0.94
 &$\circ$ &$\circ$ &$\circ$ &$\circ$ &$\times$   \\                                                            
 43   \ \dotfill \  &$    50.84$&$   -82.26$&$   5.88$&   0.33 &$  -6.38$&   0.38
 &$  70.56$&   2.68
 &$\circ$ &$\circ$ &$\circ$ &$\circ$ &$\times$   \\                                                            
 44  \ \dotfill \  &$    34.87$&$  -150.80$&$   3.10$&   0.47 &$ -10.59$&   0.51
 &$  94.26$&   3.93
 &$\times$ &$\circ$ &$\circ$ &$\circ$ &$\times$   \\                                                           
 45  \ \dotfill \  &$    38.98$&$  -157.15$&$   4.18$&   0.37 &$ -10.30$&   0.42
 &$ 105.74$&   5.20
 &$\times$ &$\times$ &$\circ$ &$\circ$ &$\circ$   \\                                                           
 46  \ \dotfill \  &$    34.95$&$  -152.89$&$  14.90$&   0.57 &$ -12.15$&   0.58
 &$ 116.29$&   1.55
 &$\circ$ &$\times$ &$\circ$ &$\circ$ &$\times$   \\                                                           
 47  \ \dotfill \  &$    52.36$&$  -175.06$&$  -1.21$&   0.64 &$ -20.59$&   0.90
 &$ 125.55$&   1.82
 &$\circ$ &$\circ$ &$\circ$ &$\times$ &$\times$   \\                                                           
 48   \ \dotfill \  &$    52.37$&$  -175.07$&$  -1.49$&   1.26 &$ -20.26$&   1.72
 &$ 128.17$&   1.96
 &$\circ$ &$\circ$ &$\circ$ &$\times$ &$\times$   \\                                                           
 49  \ \dotfill \  &$    52.06$&$  -175.05$&$   0.43$&   1.38 &$ -17.51$&   1.22
 &$ 133.98$&   3.51
 &$\circ$ &$\circ$ &$\circ$ &$\times$ &$\times$   \\                                                           
 50  \ \dotfill \  &$    52.15$&$  -175.06$&$   5.19$&   0.26 &$ -14.23$&   0.30
 &$ 135.91$&   3.58
 &$\circ$ &$\circ$ &$\circ$ &$\circ$ &$\times$   \\                                                            
\hline                                
 
\end{tabular}

\footnotemark[a]\h2o maser features detected toward \i18286. The feature is designated as \i18286:I2013-{\it N}, 
where {\it N}is the ordinal source number given in this column (I2013 stands for sources found by Imai 
\etal\ and listed in 2013).\\
\footnotemark[b]Relative value with respect to the motion of the position-reference maser feature located 
at the map origin. \\
\footnotemark[c]Relative value with respect to the local standard of rest. \\
\footnotemark[d]Mean full velocity width of the maser feature at half intensity. \\
\end{table*}

\clearpage
\begin{table*}[h]
\caption{Parameters of the best fit 3D spatio-kinematical model of the \h2o\ masers in \i18286}
\label{tab:kinematic-model}

\scriptsize
\begin{tabular}{lr@{$\pm$}l} \hline \hline
Parameter \hspace{1.5cm} & \multicolumn{2}{c}{Value} \\ \hline
$\Delta X_{0}$ [mas] \footnotemark[a] \dotfill & 24 & 3 \\
$\Delta Y_{0}$ [mas] \footnotemark[a] \dotfill & 49 & 4 \\
$V_{0X}$ [km~s$^{-1}$] \footnotemark[a] \dotfill & 13 & 2 \\
$V_{0Y}$ [km~s$^{-1}$] \footnotemark[a] \dotfill & 21 & 4 \\
$V_{0Z}$ [km~s$^{-1}$] \footnotemark[b] \dotfill & \multicolumn{2}{c}{65} \\ 
$D$ [kpc] \footnotemark[b] \  \dotfill & \multicolumn{2}{c}{3.6} \\ \hline
$\sqrt{\chi^{2}}$ \footnotemark[c] \dotfill & \multicolumn{2}{c}{4.22}  \\ \hline
\end{tabular}

\footnotemark[a]Relative value respect to the maser feature \i18286:I2013-{\it 14}. \\
\footnotemark[b]Assumed value. \\
\footnotemark[c]The $\chi^{2}$-square value was obtained in the model fitting based on the least-squares method. \\

\caption{Parameters of the fitted maser spot motion}\label{tab:astrometry}

\scriptsize
\begin{tabular}{lr@{$\pm$}l r@{$\pm$}l r@{$\pm$}l r@{$\pm$}l r@{$\pm$}l} \hline \hline
\multicolumn{1}{c}{\vlsr} & \multicolumn{2}{c}{$X_{\mbox \scriptsize 0}$\footnotemark[a]} 
& \multicolumn{2}{c}{$Y_{\mbox \scriptsize 0}$\footnotemark[a]}
& \multicolumn{2}{c}{$\mu_{X}$} & \multicolumn{2}{c}{$\mu_{Y}$} & \multicolumn{2}{c}{$\pi$} \\
\multicolumn{1}{c}{[km~s$^{-1}$]} & \multicolumn{2}{c}{[mas]} 
& \multicolumn{2}{c}{[mas]} & \multicolumn{2}{c}{[mas~yr$^{-1}$]}
& \multicolumn{2}{c}{[mas~yr$^{-1}$]} & \multicolumn{2}{c}{[mas]}  \\ \hline
\multicolumn{11}{c}{Independent fitting} \\ \hline
 53.4 & 26.45 &    1.00 &   55.56 &    0.86 & $  -3.37$ &   0.12 & $  -7.11$ &  0.10 &  0.284 &  0.058 \\
 53.0 & 27.57 &    1.20 &   56.53 &    0.90 & $  -3.50$ &   0.15 & $  -7.24$ &   0.11 &  0.229 &  0.056 \\
\hline
\multicolumn{11}{c}{Proper motion fitting (after subtracting the annual parallax\footnotemark[b]} \\ \hline
 53.4 & 25.46 &    1.60 &   55.46 &    0.98 & $  -3.26$ &   0.19 & $  -7.10$ &  0.12 &  
 \multicolumn{2}{c}{...} \\
 53.0 & 25.10 &    2.22 &   56.90 &    0.94 & $  -3.22$ &   0.27 & $  -7.28$ &   0.11 &
 \multicolumn{2}{c}{...} \\
\hline
\multicolumn{11}{c}{Combined fitting to an annual parallax and residuals of position offset and linear motion} \\ \hline
Combined & 1.56 & 0.80 & $-$0.06 & 0.63 & $  -0.17 $ &  0.10 & $ 0.01 $ &  0.08 &   0.277 & 0.041 \\
\hline
\end{tabular}

\footnotemark[a]Position at the epoch J2000.0 with respect to the delay-tracking center (see main text).\\
\footnotemark[b]The adopted annual parallax was given in the combined fitting in the converging phase of iterations. \\

\caption{Location and 3D motion of \i18286 in the Milky Way estimated from the VERA astrometry}
\label{tab:MW-motion}
\begin{tabular}{lr@{$\pm$}l@{}} \hline \hline
Parameter \hspace{3.8cm} & \multicolumn{2}{c}{Value} \\ \hline
Galactic coordinates, $(l,b)$ [deg]\footnotemark[a] \dotfill  & \multicolumn{2}{c}{(21.80, $-0.13$)} \\
Heliocentric distance, $D$ [kpc]\footnotemark[a] \dotfill  & 3.61 & 0.63 \\
Systemic LSR velocity, $V_{\rm sys}$ [km~s$^{-1}$]\footnotemark[a] \dotfill & $60$ & 5 \\
$R_{\rm 0}$ [kpc]\footnotemark[b] \dotfill  & \multicolumn{2}{c}{8.0} \\
$\Theta_{\rm 0}$ [km~s$^{-1}$]\footnotemark[b] \dotfill & \multicolumn{2}{c}{220} \\
$(U_{\odot}, V_{\odot}, W_{\odot})$ [km~s$^{-1}$]\footnotemark[c] \dotfill 
& \multicolumn{2}{c}{(7.5, 13.5, 6.8)} \\
$z_{\rm 0}$ [pc]\footnotemark[d] \dotfill & \multicolumn{2}{c}{16} \\
$R_{\rm gal}$ [kpc] \dotfill & 4.84 & 0.50 \\
$z$ [pc] \dotfill & 7 & 1 \\ \hline 
\multicolumn{3}{c}{Case 1: fixed in the system} \\ \hline
$V_{R}$ [km~s$^{-1}$]  \dotfill & 66 & 19 \\
$V_{\theta}$ [km~s$^{-1}$]  \dotfill & 148 & 25 \\
$V_{z}$ [km~s$^{-1}$]  \dotfill & $-1$ & 20 \\ \hline
\multicolumn{3}{c}{Case 2: moving in ($-0.76$, $-1.23$) [mas~yr$^{-1}$] with respect} \\
\multicolumn{3}{c}{to the system (result of the kinematical model fitting)} \\ \hline
$V_{R}$ [km~s$^{-1}$]  \dotfill & 51 & 17 \\
$V_{\theta}$ [km~s$^{-1}$]  \dotfill & 169 & 22 \\
$V_{z}$ [km~s$^{-1}$]  \dotfill & $-2$ & 20 \\ \hline
\multicolumn{3}{c}{Case 3: moving in ($-3$, $-5$) [mas~yr$^{-1}$]} \\
\multicolumn{3}{c}{with respect to the system} \\ \hline
$V_{R}$ [km~s$^{-1}$]  \dotfill & 6 & 13 \\
$V_{\theta}$ [km~s$^{-1}$]  \dotfill & 228 & 17 \\
$V_{z}$ [km~s$^{-1}$]  \dotfill & $-7$ & 20 \\ \hline
\end{tabular}

\footnotemark[a]Input value for \i18286.\\
\footnotemark[b]Input value for the Sun in the Milky Way. \\
\footnotemark[c]Motion of the Sun with respect to the local standard of rest, 
cited from \citet{fra09} (c.f., \cite{deh98}). \\
\footnotemark[d]Height of the Sun from the Galactic mid-plane, cited from \citet{ham95}. \\
\end{table*}

\end{document}